\documentclass[prd,amsmath,amssymb,superscriptaddress,preprintnumbers,twocolumn,nofootinbib,10pt]{revtex4-1}

\pdfoutput=1

\usepackage{graphicx}
\usepackage{dcolumn}
\usepackage{bm}
\usepackage{amssymb}
\usepackage{latexsym}
\usepackage{booktabs}
\usepackage{amsmath}
\usepackage{multirow}
\usepackage{url}

\usepackage{float}
\usepackage[colorlinks=true, linkcolor=red, citecolor=blue]{hyperref}

\usepackage[normalem]{ulem}
\usepackage{color}
\usepackage{array}
\usepackage{enumerate}

\begin{document}

\renewcommand{\bibfont}{\footnotesize}

\title{Measuring neutrino mass in light of ACT DR6 and DESI DR2}

\author{Lu Feng}
\affiliation{College of Physical Science and Technology, Shenyang Normal University, Shenyang 110034, China}
\affiliation{Liaoning Key Laboratory of Cosmology and Astrophysics, College of Sciences, Northeastern University, Shenyang 110819, China}
\author{Tian-Nuo Li}
\affiliation{Liaoning Key Laboratory of Cosmology and Astrophysics, College of Sciences, Northeastern University, Shenyang 110819, China}
\author{Guo-Hong Du}
\affiliation{Liaoning Key Laboratory of Cosmology and Astrophysics, College of Sciences, Northeastern University, Shenyang 110819, China}
\author{Jing-Fei Zhang}
\affiliation{Liaoning Key Laboratory of Cosmology and Astrophysics, College of Sciences, Northeastern University, Shenyang 110819, China}
\author{Xin Zhang}\thanks{Corresponding author}
\email{zhangxin@neu.edu.cn}
\affiliation{Liaoning Key Laboratory of Cosmology and Astrophysics, College of Sciences, Northeastern University, Shenyang 110819, China}
\affiliation{MOE Key Laboratory of Data Analytics and Optimization for Smart Industry, Northeastern University, Shenyang 110819, China}
\affiliation{National Frontiers Science Center for Industrial Intelligence and Systems Optimization, Northeastern University, Shenyang 110819, China}

\begin{abstract}
The recent release of high-precision cosmological data, particularly the small-scale cosmic microwave background (CMB) measurements from ACT and baryon acoustic oscillation (BAO) data from DESI, has opened a new landscape for probing the neutrino mass.
In this work, we present updated constraints on the total neutrino mass, $\sum m_\nu$, and its hierarchy within the $\Lambda$CDM, $w$CDM, holographic dark energy (HDE), and $w_0w_a$CDM models, using the latest ACT DR6, DESI DR2, and DESY5 datasets.
We find that the upper limits on $\sum m_\nu$ are critically governed by the evolutionary behavior of the dark energy equation of state. Specifically, models exhibiting early-time quintessence features (e.g., HDE) yield the most stringent constraints, whereas those allowing for early-time phantom behavior (e.g., $w_0w_a$CDM) result in significantly looser bounds.
Despite these model-dependent variations, we observe a robust hierarchy dependence across all scenarios, where the inverted hierarchy consistently yields weaker constraints and the degenerate hierarchy consistently yields tightest constraints.
Our analysis demonstrates that the improved small-scale CMB information from ACT, combined with high-precision BAO data, systematically tightens the limits on $\sum m_\nu$, providing a crucial benchmark for future neutrino mass measurement.

\end{abstract}

\maketitle

\section{Introduction}
The standard cosmological model, $\Lambda$ cold dark matter ($\Lambda$CDM), has provided an excellent fit to cosmological observations since the discovery of cosmic acceleration \cite{SupernovaSearchTeam:1998fmf,SupernovaCosmologyProject:1998vns}. However, it still encounters two major theoretical challenges, namely the ``cosmic coincidence'' and ``fine-tuning'' problems \cite{Sahni:1999gb,Carroll:2000fy,Weinberg:2000yb,Frieman:2008sn}. In addition, it is subject to several emerging observational tensions, such as the Hubble constant ($H_0$) tension and the matter fluctuation amplitude ($S_8$) discrepancy \cite{Vagnozzi:2019ezj,Vagnozzi:2023nrq,Vagnozzi:2021tjv,Verde:2019ivm,DiValentino:2020zio,DiValentino:2020vvd,DiValentino:2021izs,Vagnozzi:2021gjh,Abdalla:2022yfr,CosmoVerseNetwork:2025alb}. To address these issues, various dark energy (DE) models~\cite{Linder:2002et,Peebles:2002gy,Zhang:2004gc,Zhang:2005rj,Zhang:2005yz,Zhang:2005hs,Copeland:2006wr,Wang:2006qw,Zhang:2007sh,Zhang:2009un,Zhang:2012uu,Costa:2013sva,Bull:2015stt,Feng:2016djj,Wang:2016och,Feng:2017mfs,Feng:2017usu,Feng:2018yew,Pan:2019gop,DiValentino:2020kpf,Perivolaropoulos:2021jda,Feng:2021ipq,Giare:2024smz,Wang:2024dka,Huang:2025som,Shah:2025ayl,Pang:2025lvh,You:2025uon,Goswami:2025uih,Barua:2025ypw,Li:2025ops,Qiang:2025cxp,Feng:2025wbz,vanderWesthuizen:2025rip,Goh:2025upc,Alam:2025epg,Li:2025vqt,Zhu:2025lrk} have been proposed and extensively studied as potential alternatives. 
Apart from DE, massive neutrinos constitute another fundamental component of the universe and are routinely considered in the interpretation of cosmological observations.

From the particle physics perspective, solar and atmospheric neutrino experiments have demonstrated that neutrinos are massive and exhibit significant flavor mixing (see Ref.~\cite{Lesgourgues:2006nd} for a review). These measurements have established two independent mass-squared splittings: solar and reactor experiments determine $\Delta m_{21}^2 \simeq 7.5\times10^{-5}~\mathrm{eV}^2$, while atmospheric and accelerator-based experiments yield $|\Delta m_{31}^2| \simeq 2.5\times10^{-3}~\mathrm{eV}^2$. Such results imply that the neutrino mass can follow either the normal hierarchy (NH), $m_1 < m_2 \ll m_3$, or the inverted hierarchy (IH), $m_3 \ll m_1 < m_2$, where $m_1$, $m_2$, and $m_3$ denote the masses of the three mass eigenstates. Despite the precise determination of these mass-squared differences, the absolute neutrino mass scale and the hierarchy of the mass eigenstates remain unresolved, representing major outstanding questions in contemporary particle physics and cosmology. Given that neutrinos affect the growth of cosmic structures and the anisotropies in the cosmic microwave background (CMB), cosmological observations offer a powerful complementary approach to constraining neutrino masses.

In fact, cosmological constraints on the total neutrino mass have been extensively investigated (see, e.g., Refs. \cite{Zhang:2015rha,Geng:2015haa,Chen:2015oga,Allison:2015qca,Zhang:2015uhk,Cuesta:2015iho,Huang:2015wrx,Chen:2016eyp,Moresco:2016nqq,Giusarma:2016phn,Lu:2016hsd,Wang:2016tsz,Zhao:2016ecj,Kumar:2016zpg,Xu:2016ddc,Vagnozzi:2017ovm,Guo:2017hea,Zhang:2017rbg,Li:2017iur,Yang:2017amu,Lorenz:2017fgo,Wang:2017htc,Zhao:2017jma,Vagnozzi:2018jhn,Wang:2018lun,Giusarma:2018jei,Guo:2018gyo,Zhao:2018fjj,Loureiro:2018pdz,Feng:2019mym,Zhang:2019ipd,Feng:2019jqa,Liu:2020vgn,Yang:2020tax,Zhang:2020mox,Li:2020gtk,Yang:2020ope,Jin:2022tdf,Reeves:2022aoi,Tanseri:2022zfe,Pang:2023joc,Feng:2024lzh,Du:2024pai,Elbers:2025vlz,Zhou:2025nkb,Ivanov:2026dvl,FrancoAbellan:2026ori,Ladeira:2026jne,Nair:2025uyn,Chudaykin:2025lww,DOnofrio:2025cuk,Sharma:2025iux,Barua:2025adv,Du:2025xes,Giare:2025ath,DESI:2025ffm,RoyChoudhury:2025dhe,RoyChoudhury:2024wri,Jiang:2024viw}). 
In particular, analyses within the framework of dynamical DE models (e.g., Refs. \cite{Zhang:2015uhk,Zhang:2017rbg,Li:2017iur,Yang:2017amu,Vagnozzi:2018jhn,Zhao:2018fjj,Zhang:2020mox,Du:2024pai,Zhou:2025nkb}) suggest that the properties of DE can play a significant role in the measurement of the total neutrino mass.
In this context, cosmological constraints on the neutrino mass are intrinsically model-dependent and can vary significantly with the assumed DE properties. As a result, the presence of dynamical DE introduces additional degeneracies and uncertainties in neutrino mass measurements, limiting the robustness of current constraints. Breaking these degeneracies and achieving reliable neutrino mass bounds therefore requires both a systematic exploration of dynamical DE models and high-precision cosmological observations.

Recently, the Atacama Cosmology Telescope (ACT) collaboration reported their final measurements of the CMB power spectrum, along with the corresponding cosmological implications~\cite{AtacamaCosmologyTelescope:2025blo,AtacamaCosmologyTelescope:2025nti}, based on five years of observations from Data Release 6 (DR6).
Since the ACT results provide measurements of the small-scale region of the CMB spectrum, they play an essential role in advancing the study of cosmological problems~\cite{AtacamaCosmologyTelescope:2025nti,DESI:2025gwf}.
In particular, they provide enhanced sensitivity to the imprints of massive neutrinos on the CMB, allowing for tighter constraints on neutrino mass~\cite{AtacamaCosmologyTelescope:2025nti}.

In addition to the early universe information provided by ACT through CMB measurements, complementary constraints on the late-time cosmic expansion history can be obtained from large-scale structure observations.
In this context, baryon acoustic oscillation (BAO) measurements play a central role, and the Dark Energy Spectroscopic Instrument (DESI) collaboration has provided substantial evidence for a dynamical DE component~\cite{DESI:2024mwx}, based on BAO measurements using galaxies, quasars, and Lyman-$\alpha$ forest tracers from the first data release (DR1) of DESI~\cite{DESI:2024uvr,DESI:2024lzq}.
Interestingly, this dynamical DE evidence is further supported~\cite{DESI:2025zgx,DESI:2025fii} by the DESI second data release (DR2), which includes more than 14 million galaxies and quasars, based on three years of operation~\cite{DESI:2025zpo}. Beyond the evidence for dynamical DE, DESI BAO data have also been widely used in various cosmological analyses~\cite{Li:2024qso,Li:2024qus,Li:2025owk,Du:2025iow,Feng:2025mlo,Silva:2025hxw,Wang:2025ljj,Yang:2025ume,Ling:2025lmw,Wang:2025dtk,Li:2025eqh,Li:2025dwz,Qiang:2025cxp,Li:2025htp,Wang:2025vtw,Yang:2025oax,Petri:2025swg,Plaza:2025nip,Wu:2025vfs,Jia:2025poj,Liu:2025evk,Pedrotti:2025ccw,Reeves:2025xau,Xu:2025nsn,Li:2025muv,Zhang:2025dwu,Blanco:2025vva,Yao:2025twv,Wang:2025xvi,Du:2025csv,Zhang:2025bmk,Li:2025vqt,Song:2025bio,Carloni:2025dqt,Wang:2025djw,Yang:2025gaz,Li:2025vuh,Krolewski:2025deb,Zhang:2025lam,Yadav:2025wbc,Capozziello:2025qmh,Zhao:2025rya,Ghedini:2025epp,deCruzPerez:2025dni,Paliathanasis:2026ymi,Figueruelo:2026eis,Wang:2026kbg,Wang:2026lno,Li:2026xaz,Schiavone:2026agq,Yin:2026gss,Aboubrahim:2026tks,Millard:2026wnd,Du:2026cly,Fazzari:2025lzd,Giare:2025pzu,Giare:2024gpk,Giare:2024oil,Pan:2025qwy}.

In this study, by combining the latest cosmological observations, including ACT DR6 CMB data, DESI DR2 BAO measurements, and DESY5 type Ia supernova (SN) data, we aim to provide a novel measurement of the neutrino mass from a cosmological perspective. To perform a comprehensive and systematic analysis, we consider a variety of DE models, including $\Lambda$CDM, $w$CDM, holographic DE (HDE), and the $w_0w_a$CDM model, while explicitly accounting for the possible neutrino mass hierarchies. Our analysis provides critical insights into how the neutrino mass hierarchy and the nature of DE influence neutrino mass measurements, revealing how parameter degeneracies under different cosmological backgrounds affect the robustness of the mass upper limits.

This work is organized as follows. In Sec.~\ref{sec2}, the $\Lambda$CDM, $w$CDM, HDE, and $w_0w_a$CDM models are briefly described. In Sec.~\ref{sec3}, we introduce the observational data and the analysis method used in this paper.  In Sec.~\ref{sec4}, we give the constraint results and provide some relevant discussions. Finally, the conclusion is given in Sec.~\ref{sec5}.

\section{Dark energy models}\label{sec2}
In this work, we take into account the mass-squared splittings among the three active neutrino mass eigenstates. The total neutrino mass is therefore given by
\begin{equation}
    \sum m_\nu = m_1 + \sqrt{m_1^2 + \Delta m_{21}^2} + \sqrt{m_1^2 + |\Delta m_{31}^2|},
\end{equation}
for the NH, and
\begin{equation}
    \sum m_\nu = m_3 + \sqrt{m_3^2 + |\Delta m_{31}^2| + \Delta m_{21}^2} + \sqrt{m_3^2 + |\Delta m_{31}^2|},
\end{equation}
for the IH.
For the degenerate hierarchy (DH), we adopt the approximation $\sum m_\nu = 3m$.
In our analyses, we impose the priors, $\sum m_\nu > 0.06~\mathrm{eV}$ for NH, $\sum m_\nu > 0.10~\mathrm{eV}$ for IH and $\sum m_\nu > 0~\mathrm{eV}$ for DH.

To investigate the impact of DE on the constraints of the total neutrino mass, we consider four representative DE models. A brief overview of these models is provided below.

\subsection{The $\Lambda$CDM model}
The $\Lambda$CDM model includes the vacuum energy component that acts as DE with an equation of state (EoS) $w=-1$. The six base parameters of this model include the energy densities of baryons $\omega_\mathrm{b}\equiv \Omega_\mathrm{b} h^2$ and cold dark matter, $\omega_\mathrm{c}\equiv \Omega_\mathrm{c} h^2$, the acoustic angular scale $\Theta_{\rm{s}}$, the optical depth of reionization $\tau$, the power law spectral index $n_\mathrm{s}$, and the amplitude of initial curvature perturbation $A_\mathrm{s}$. 

When the total neutrino mass is included in the $\Lambda$CDM model, the model considered in this paper is called the $\Lambda$CDM+$\sum m_\nu$ model. Thus, there are seven independent parameters in total for $\Lambda$CDM+$\sum m_\nu$ and the parameter space vector is
\begin{equation}
{\bf P_1}=\{\omega_\mathrm{b}, \omega_\mathrm{c}, \Theta_{\rm{s}}, \tau, n_\mathrm{s}, \ln (10^{10}A_\mathrm{s}), \sum m_\nu\}.
\end{equation}

\subsection{The $w$CDM model}
The $w$CDM model assumes that the EoS parameter of DE is constant. This model has one more parameter $w$ compared to $\Lambda$CDM.

When we include the total neutrino mass in the $w$CDM model, the model considered in this paper is called the $w$CDM+$\sum m_\nu$ model. Thus, there are eight independent parameters in total for $w$CDM+$\sum m_\nu$ and the parameter space vector is
\begin{equation}
{\bf P_2}=\{\omega_\mathrm{b}, \omega_\mathrm{c}, \Theta_{\rm{s}}, \tau, n_\mathrm{s}, \ln (10^{10}A_\mathrm{s}), w, \sum m_\nu\}.
\end{equation}


\begin{table*}[!htbp]
\renewcommand\arraystretch{1.5}
\centering
\caption{The 1$\sigma$ or 2$\sigma$ confidence ranges of cosmological parameters obtained from the CMB+BAO+SN for the $\Lambda$CDM, $w$CDM, HDE and $w_0w_a$CDM models. For the parameter $\sum m_\nu$, the 2$\sigma$ upper limits are reported. For all other parameters, the 1$\sigma$ confidence intervals are reported.}
\footnotesize
\setlength{\tabcolsep}{5pt}
\label{tab}
\resizebox{0.9\textwidth}{!}
{%
\begin{tabular}{lcccccc}
\hline
Parameter  & $H_0$ [${\rm km}\ {\rm s^{-1}}\  {\rm Mpc^{-1}}$] & $\Omega_{\rm m}$ & $w\ {\rm or}\ w_0$ & $c$ & $w_a$ & $\sum m_\nu\,[\mathrm{eV}]$
 \\
\hline
\multicolumn{7}{l}{$\boldsymbol{\Lambda{\rm CDM}}$} \\
DH    & $68.39\pm 0.30$  & $0.3023\pm 0.0038$ & --- & --- & --- & $< 0.071$\\
NH    & $68.18\pm 0.30$  & $0.3042\pm 0.0039$ & --- & --- & --- & $< 0.121$\\
IH    & $68.04\pm 0.29$  & $0.3053\pm 0.0038$ & --- & --- & --- & $< 0.148$\\
\hline
\multicolumn{7}{l}{$\boldsymbol{w{\rm CDM}}$} \\
DH    & $67.43\pm 0.55$ & $0.3085\pm 0.0049$ & $-0.954\pm 0.022$ & --- & --- & $< 0.057$\\
NH    & $67.43\pm 0.54$ & $0.3092\pm 0.0049$ & $-0.964\pm 0.022$ & --- & --- & $< 0.110$ \\
IH    & $67.43\pm 0.54$ & $0.3095\pm 0.0049$ & $-0.971\pm 0.022$ & --- & --- & $< 0.143$ \\
\hline
\multicolumn{7}{l}{$\boldsymbol{{\rm HDE}}$} \\
DH    & $67.39\pm 0.53$  & $0.3009\pm 0.0048$ & --- & $0.735\pm 0.024$ & --- & $< 0.027$\\
NH    & $67.33\pm 0.54$  & $0.3020\pm 0.0049$ & --- & $0.724\pm 0.024$ & --- & $< 0.087$\\
IH    & $67.31\pm 0.54$  & $0.3025\pm 0.0049$ & --- & $0.716\pm 0.023$ & --- & $< 0.123$\\
\hline
\multicolumn{7}{l}{$\boldsymbol{w_0w_a{\rm CDM}}$} \\
DH    & $66.95\pm 0.57$ & $0.3185\pm 0.0057$ & $-0.753\pm 0.059$ & --- & $-0.86^{+0.27}_{-0.23}$ & $ < 0.131 $\\
NH    & $66.91\pm 0.57$ & $0.3197\pm 0.0058$ & $-0.740\pm 0.060$ & --- & $-0.95^{+0.26}_{-0.23}$ & $ < 0.164 $\\
IH    & $66.86\pm 0.56$ & $0.3206\pm 0.0057$ & $-0.729\pm 0.060$ & --- & $-1.02^{+0.27}_{-0.23}$ & $ < 0.184 $\\
\hline
\end{tabular}
}
\end{table*}


\subsection{The HDE model}
The HDE model is a dynamical DE model based on the holographic principle of quantum gravity and effective quantum field theory. Based on a consideration of the holographic principle, the total energy of a system of size $L$ should not exceed the mass of a black hole with the same size \cite{Cohen:1998zx}. For the HDE model, the energy density of DE is $\rho_{\rm {de}} = 3 c^2 M^2_{\rm {Pl}} R^{-2}_{\rm EH}$, where $c$ is a dimensionless model parameter, $M_{\rm {Pl}}$ is the reduced Planck mass, and $R_{\rm EH}$ is the event horizon size of the universe. In the HDE model \cite{Li:2004rb}, the EoS is $w = - \frac{1}{3} - \frac{2}{3} \frac{\sqrt{\Omega_{\rm{de}}}}{c}$. This model has one more parameter $c$ compared to $\Lambda$CDM.

When the total neutrino mass is included in the HDE model, the model considered in this paper is called the HDE+$\sum m_\nu$ model. Thus, there are eight independent parameters in total for HDE+$\sum m_\nu$ and the parameter space vector is
\begin{equation}
{\bf P_3}=\{\omega_\mathrm{b}, \omega_\mathrm{c}, \Theta_{\rm{s}}, \tau, n_\mathrm{s}, \ln (10^{10}A_\mathrm{s}), c, \sum m_\nu\}.
\end{equation}

\subsection{The $w_0w_a$CDM model}
The $w_0w_a$CDM model can test a time-varying EoS by adopting the parametrization of $w(a) = w_0 + w_a (1 - a)$ \cite{Chevallier:2000qy,Linder:2002et}. Compared with the $\Lambda$CDM model, this model introduces two additional parameters $w_0$ and $w_a$.

When the total neutrino mass is included in the $w_0w_a$CDM model, the model considered in this paper is called the $w_0w_a$CDM+$\sum m_\nu$ model. Thus, there are nine independent parameters in total for $w_0w_a$CDM+$\sum m_\nu$ and the parameter space vector is
\begin{equation}
{\bf P_4}=\{\omega_\mathrm{b}, \omega_\mathrm{c}, \Theta_{\rm{s}}, \tau, n_\mathrm{s}, \ln (10^{10}A_\mathrm{s}), w_0, w_a, \sum m_\nu\}.
\end{equation}

It should be noted that many other DE scenarios have been proposed in the literature \cite{Copeland:2006wr,Bull:2015stt,Perivolaropoulos:2021jda,Wang:2016lxa}, for example, interacting DE (IDE) models \cite{Wang:2016lxa}. In this work, we focus on four representative DE models, which capture the main phenomenological features of DE and are sufficient to explore the impact of DE properties on cosmological neutrino mass constraints.

\begin{figure*}[!htbp]
\includegraphics[scale=0.45]{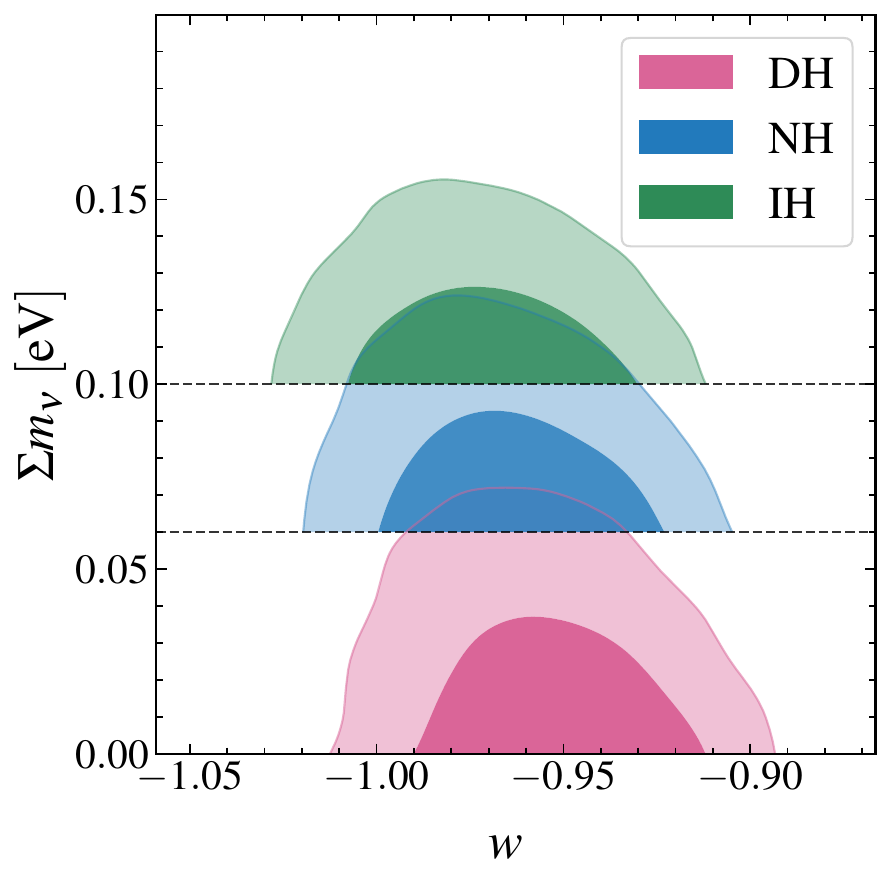}
\includegraphics[scale=0.45]{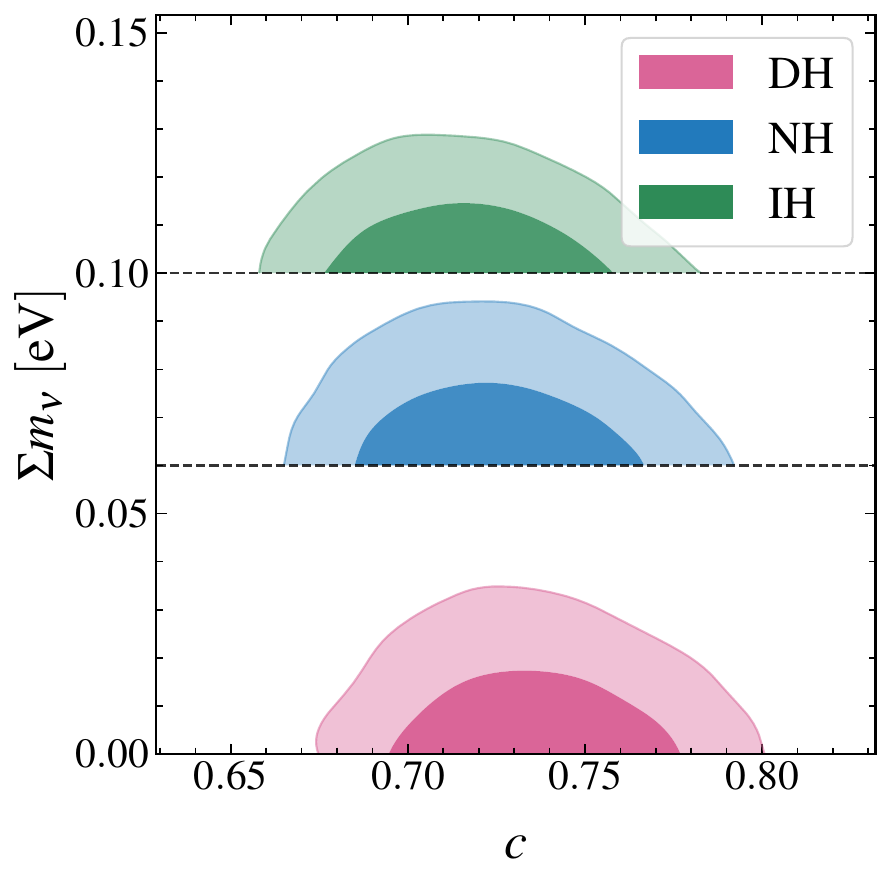}
\includegraphics[scale=0.6]{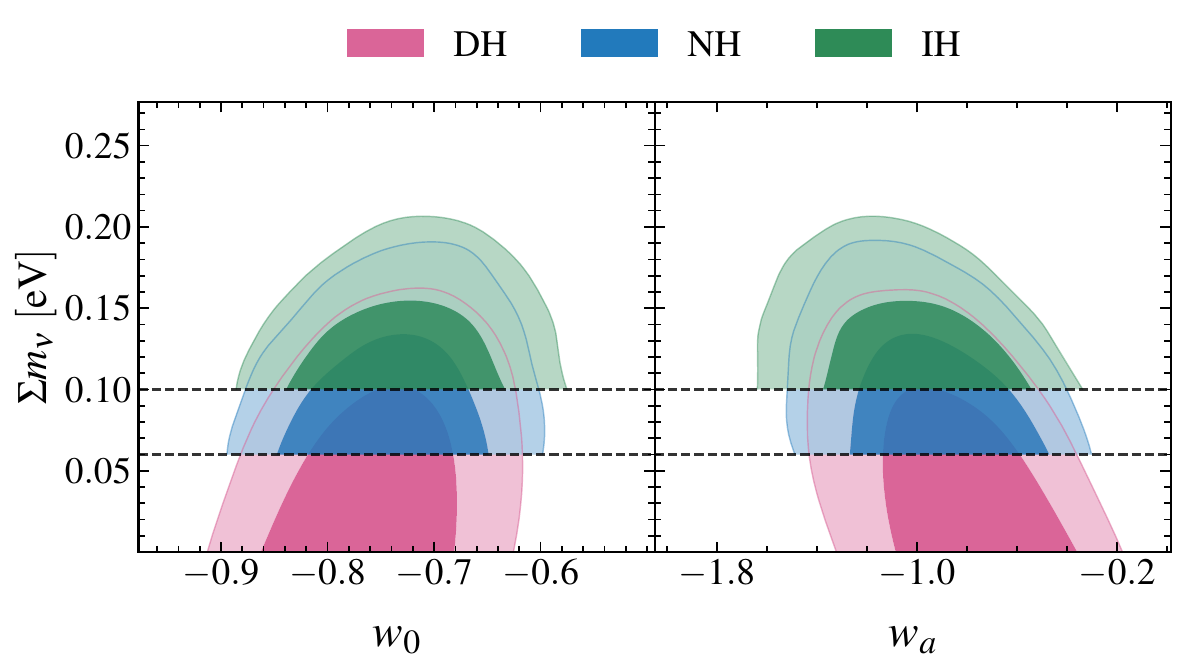}
\centering
\caption{\label{fig1} The $1\sigma$ and $2\sigma$ credible-interval contours showing the correlations between $\sum m_\nu$ and the DE parameters ($w$, $c$, $w_0$, and $w_a$) for the $w$CDM, HDE, and $w_0w_a$CDM models in the DH, NH, and IH cases, obtained from the CMB+BAO+SN data.}
\end{figure*}

\section{Data and method}\label{sec3}
In this work, we compute the theoretical predictions for these cosmological models using the \texttt{CAMB} code \cite{Lewis:1999bs}, and perform the Markov Chain Monte Carlo (MCMC) analyses using the publicly available package \texttt{Cobaya} \cite{Torrado:2020dgo,Lewis:2002ah,Lewis:2013hha}.  The convergence of the MCMC chains is assessed using the Gelman--Rubin statistic $R-1<0.02$ \cite{Gelman:1992zz} and the MCMC chains are analyzed using the public package \texttt{GetDist} \cite{Lewis:2019xzd}. In our main analysis, we perform a joint analysis of the following datasets:

{\it The CMB data}: We use the temperature and polarization likelihoods from ACT DR6 \cite{AtacamaCosmologyTelescope:2025blo}. Furthermore, we also use the CMB lensing data from ACT DR6 \cite{ACT:2023kun,ACT:2023dou} and Planck NPIPE PR4 data \cite{Carron:2022eyg,Carron:2022eum}.

{\it The BAO data}: We utilize the BAO observational data from DESI DR2, which includes observations of the bright galaxy sample, luminous red galaxies, emission line galaxies, quasars, as well as the Lyman-$\alpha$ forest. These measurements are detailed in Table IV of Ref. \cite{DESI:2025zgx}.

{\it The SN data}: We use the SN data from DESY5. This dataset consists of 194 low-redshift supernovae (SNe) spanning $0.025 < z < 0.1$ complemented by 1635 SNe distributed in $0.1 < z < 1.3$~\cite{DES:2024jxu}, bringing the total number of SNe to 1829.

Note that to probe the neutrino mass, the measurements of growth of structure (such as weak lensing and redshift space distortions) are rather important. However, we do not include growth of structure measurements in this work due to unresolved systematics that may bias the inferred neutrino mass.
In what follows, we will place constraints on the dynamical DE models involving massive neutrinos, according to the constraints of the conservative CMB+BAO+SN data combination.

\section{Results and discussion}\label{sec4}

\begin{figure*}[htbp]
\includegraphics[scale=0.37]{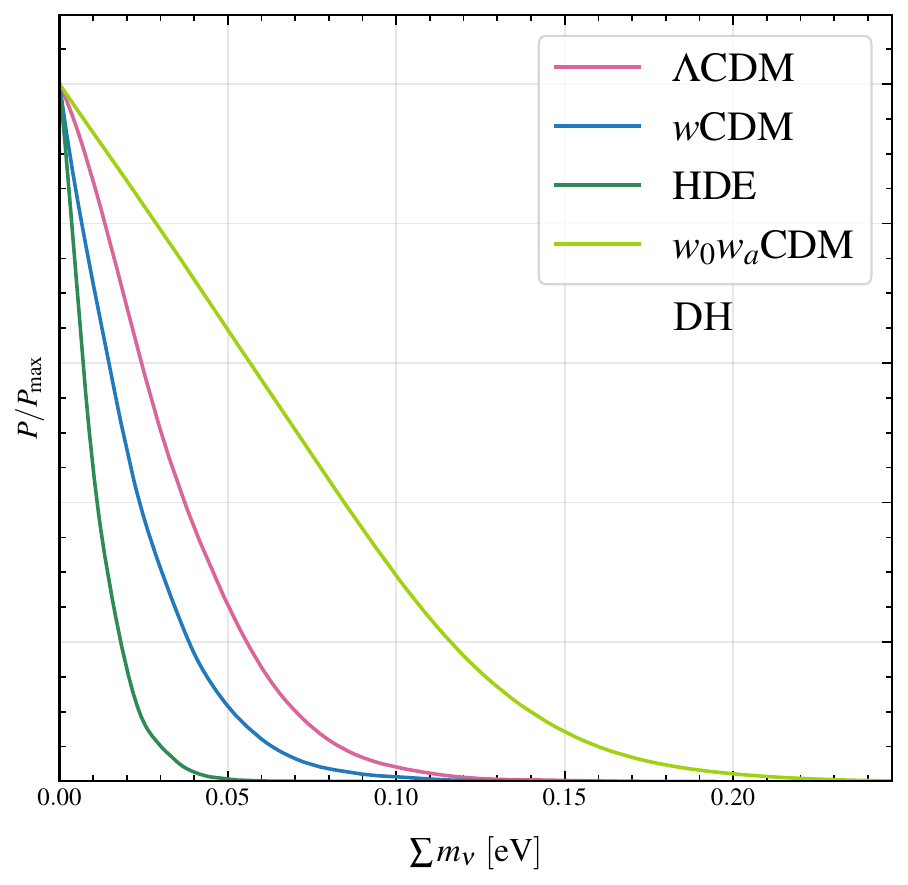}
\includegraphics[scale=0.37]{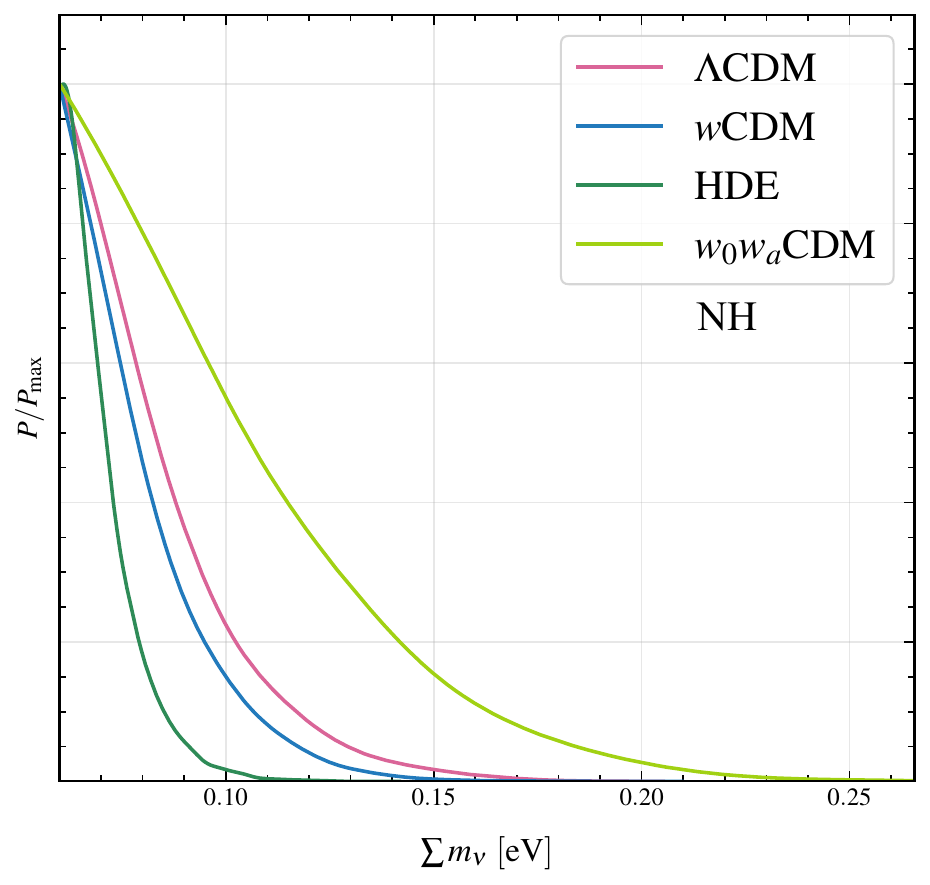}
\includegraphics[scale=0.37]{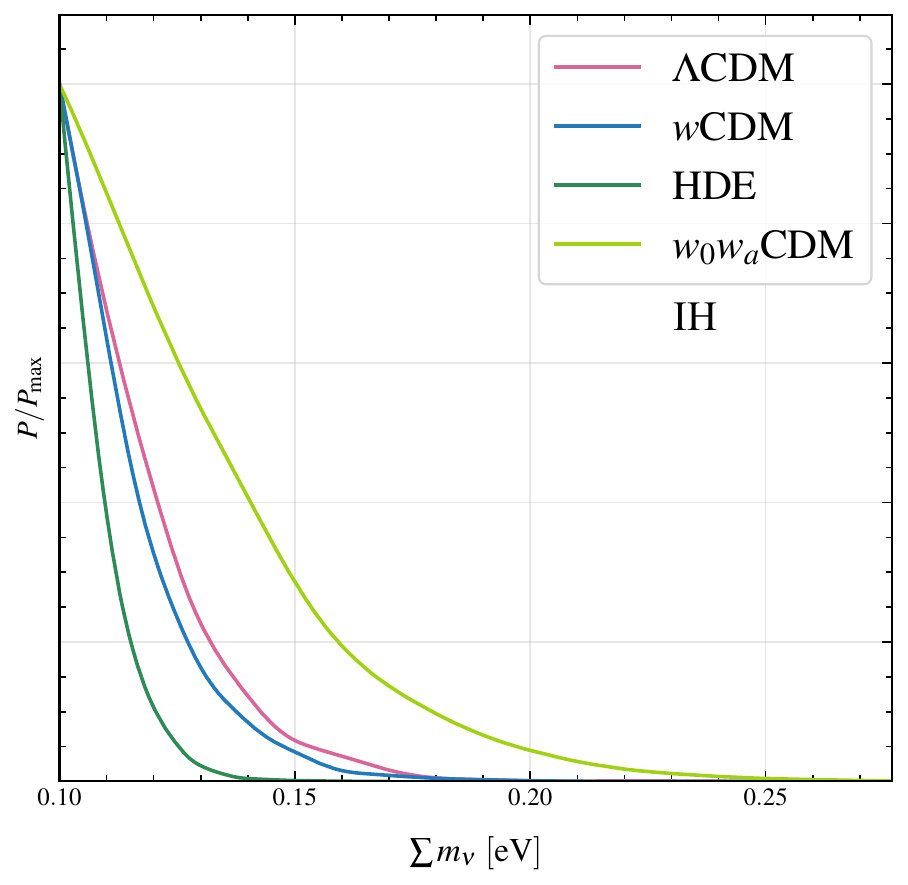}
\centering
\caption{\label{fig2} The one-dimensional marginalized posterior constraints on $\sum m_\nu$ using the CMB+BAO+SN datasets in different DE models and neutrino hierarchies.}
\end{figure*}

\begin{figure*}[htbp]
\includegraphics[scale=0.6]{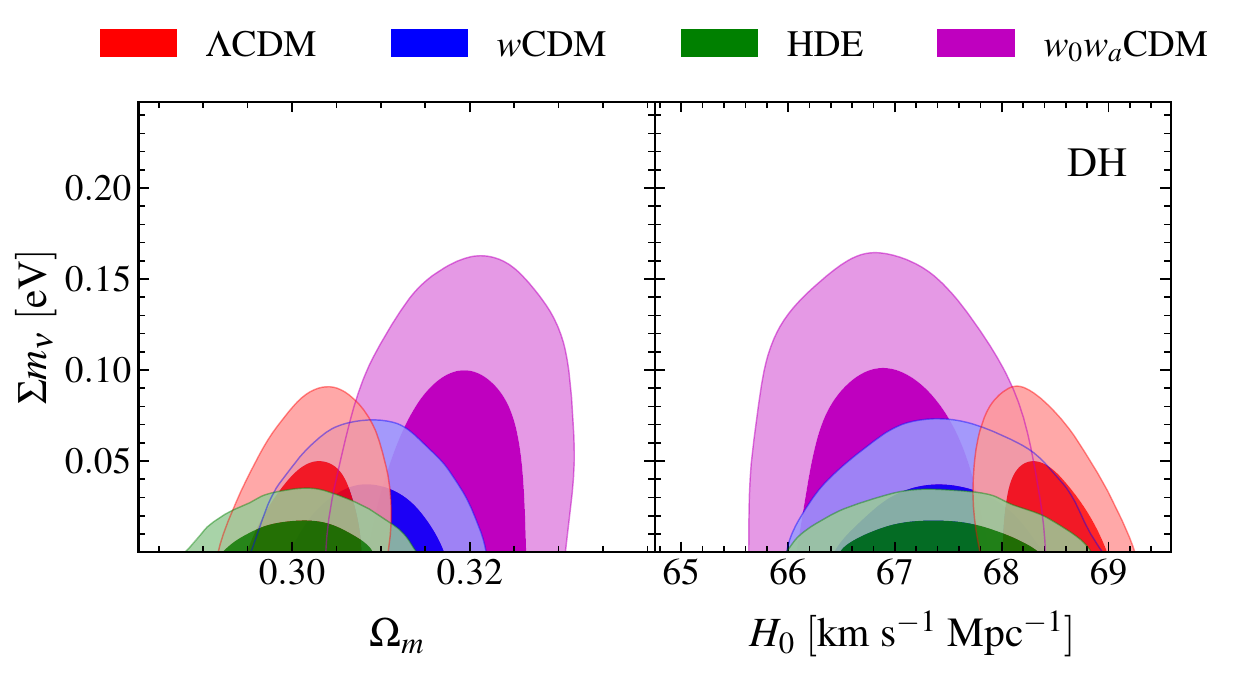}
\includegraphics[scale=0.6]{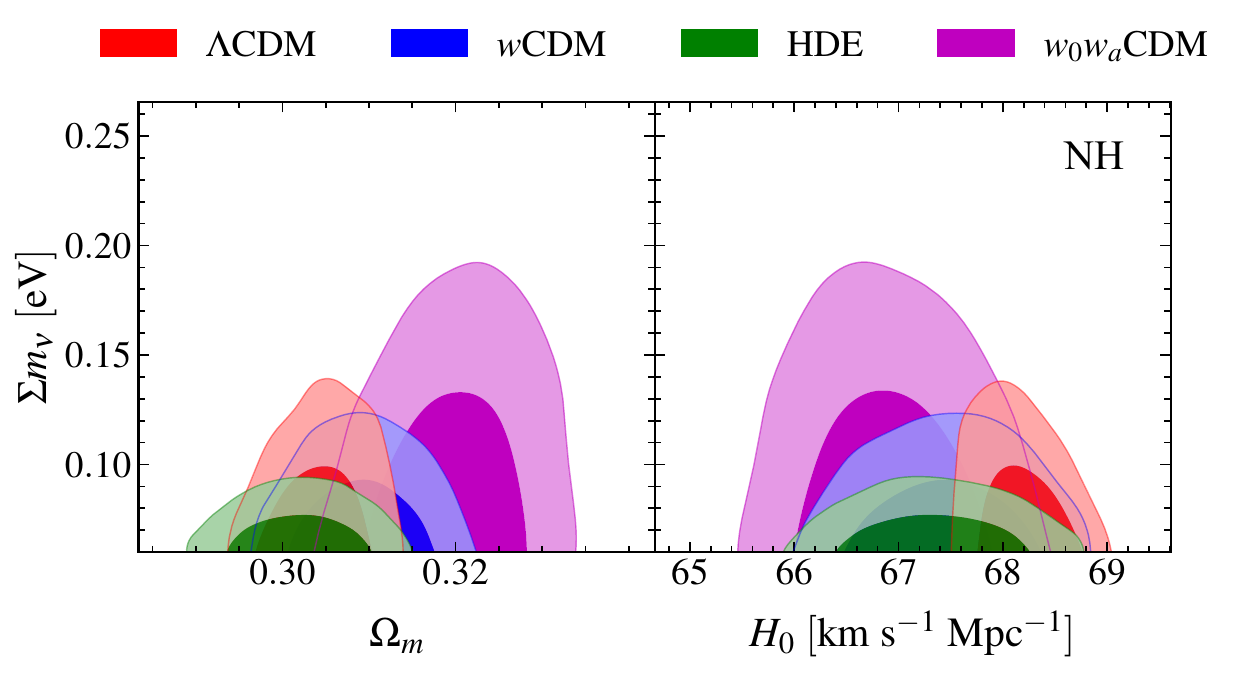}
\includegraphics[scale=0.6]{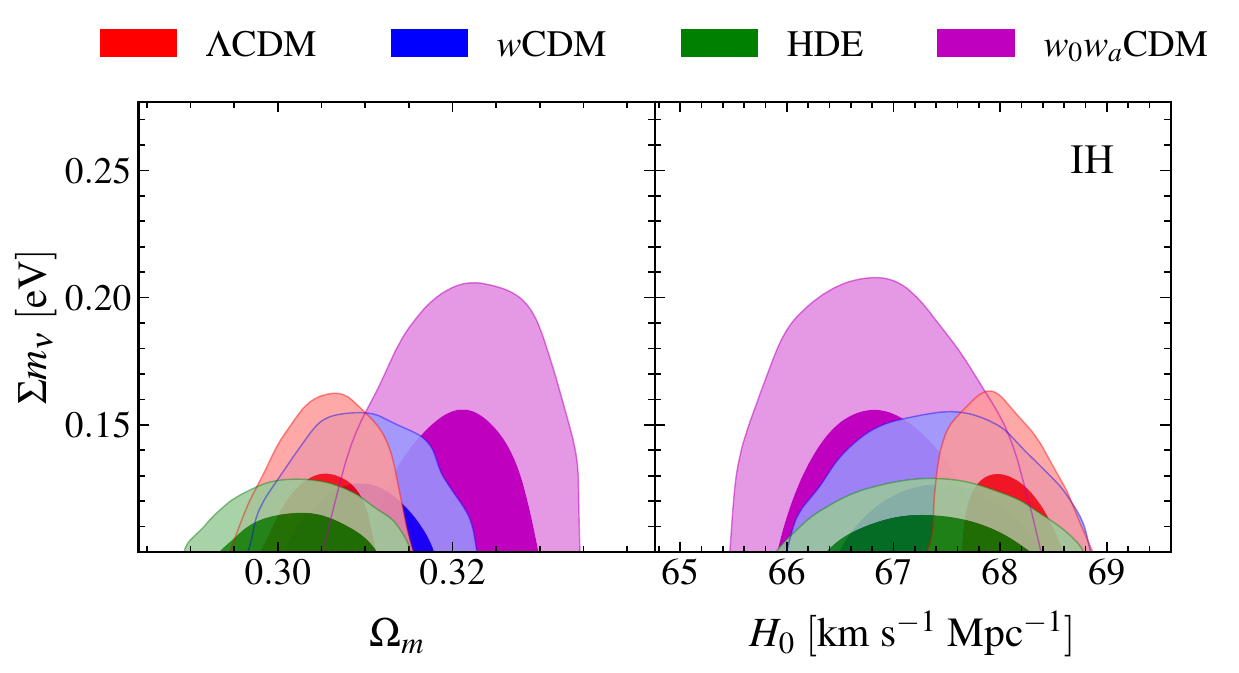}
\centering
\caption{\label{fig3} The $1\sigma$ and $2\sigma$ credible-interval contours of $\sum m_\nu$, $\Omega_{\rm m}$ and $H_0$ for the $\Lambda$CDM, $w$CDM, HDE and $w_0w_a{\rm CDM}$ models in the DH, NH, and IH cases, derived from the CMB+BAO+SN data.}
\end{figure*}

In this section, we report the constraint results of cosmological parameters for the $\Lambda$CDM, $w$CDM, HDE, and $w_0w_a$CDM models. For these models, we further consider three neutrino mass hierarchies, i.e., the NH, IH, and DH cases. We use the CMB+BAO+SN data combination to constrain these models, and the fitting results are given in Table~\ref{tab} as well as Figs.~\ref{fig1}--\ref{fig3}.
In Table~\ref{tab}, the upper limits for the sum of neutrino masses, $\sum m_\nu$, are reported at the $2\sigma$ confidence level, while constraints on other parameters are provided at the $1\sigma$ confidence level.

Across all the DE models considered, we find that the constraints on the total neutrino mass $\sum m_\nu$ exhibit a clear hierarchy dependence. In particular, the inverted hierarchy (IH) consistently yields the loosest upper limits, whereas the degenerate hierarchy (DH) leads to the tightest constraints. While the upper limits on the total neutrino mass depend sensitively on the assumed DE model, the relative hierarchy dependence of these constraints remains remarkably stable against variations in both cosmological models and observational datasets. This hierarchy dependence is in good agreement with previous studies~\cite{Wang:2016tsz,Guo:2018gyo,Feng:2019mym,Zhang:2020mox,Li:2020gtk,Du:2024pai,Zhou:2025nkb}.
Figure~\ref{fig1} provides a visual illustration of this hierarchy dependence through the two-dimensional marginalized posterior distributions, including the $\sum m_\nu$--$w$ plane in the $w$CDM model, the $\sum m_\nu$--$c$ plane in the HDE model, and the $\sum m_\nu$--$w_0$ and $\sum m_\nu$--$w_a$ planes in the $w_0w_a$CDM model, all derived from the combined CMB+BAO+SN data.
This robustness highlights the crucial role of precise small-scale CMB measurements, such as those provided by ACT DR6, when combined with high-precision BAO observations, in assessing the cosmological implications of neutrino mass and DE.

Beyond the hierarchy dependence, we find that the nature of DE, and particularly the evolutionary behavior of its EoS, plays a pivotal role in determining the cosmological constraints on the total neutrino mass. Our previous systematic studies have revealed a specific degeneracy between the DE EoS and neutrino mass measurements~\cite{Zhang:2015uhk,Zhao:2016ecj,Zhang:2017rbg,Zhang:2020mox,Du:2024pai}: a ``quintessence-like'' DE ($w > -1$) typically leads to a lower upper limit on $\sum m_\nu$ compared to $\Lambda$CDM, whereas a ``phantom-like'' DE ($w < -1$) tends to favor a larger neutrino mass. For ``quintom'' DE, where the EoS crosses $-1$, the impact on the neutrino mass depends primarily on its evolutionary behavior at early times. This degeneracy behavior provides a compelling explanation for the constraint results presented in Table~\ref{tab}. In the $w$CDM model, the fitted EoS parameter is $w \approx -0.96$ (see Table~\ref{tab}), exhibiting characteristic quintessence features. As expected, this quintessence behavior leads to tighter upper limits on the neutrino mass compared to the $\Lambda$CDM model. For instance, in the DH case, we obtain $\sum m_\nu < 0.057$ eV, compared to $< 0.071$ eV in $\Lambda$CDM. In the HDE model, although the value $c \approx 0.73 < 1$ implies that DE behaves as phantom at late times, its EoS approaches $w \to -1/3$ at early times (high redshifts), exhibiting quintessence properties. Driven by this early-time quintessence behavior, the HDE model yields the most stringent constraints on the neutrino mass, providing an upper limit of $\sum m_\nu < 0.027$ eV in the DH case. In contrast, the $w_0w_a$CDM model yields the loosest constraints. This is attributed to the fact that the EoS evolution in this model exhibits significant phantom features at early times. This early-time phantom behavior enlarges the allowable parameter space for $\sum m_\nu$, resulting in significantly relaxed neutrino mass bounds (with $\sum m_\nu < 0.131$ eV in the DH case). This result is clearly reflected in Figs.~\ref{fig2}--\ref{fig3}, which visually illustrate how the evolutionary history of the DE EoS tightens or relaxes the neutrino mass limits through parameter degeneracy.

Our conclusions are consistent with those reported in Ref.~\cite{Du:2024pai}, where the cosmological constraints on the total neutrino mass, including the effects of mass hierarchy, were investigated within the $\Lambda$CDM, $w$CDM, and $w_0w_a$CDM models using the CMB (Planck), DESI DR1 BAO, and DESY5 SN data combination. 
While the overall structure of the observational data combination remains CMB+BAO+SN, the replacement of Planck CMB data with ACT DR6 measurements represents a substantial improvement in the small-scale CMB information relevant to neutrino mass constraints.
In particular, the enhanced sensitivity of ACT DR6 to high-$\ell$ temperature and polarization anisotropies provides stronger leverage on the scale-dependent effects induced by massive neutrinos.
Compared with Ref.~\cite{Du:2024pai}, we find that the combination of ACT DR6 CMB measurements and improved DESI DR2 BAO data systematically tighten the constraints on $\sum m_\nu$ in all mass hierarchies and DE models.
This demonstrates that the latest generation of CMB and BAO observations significantly enhances the cosmological capability to probe the absolute neutrino mass scale.
Beyond the quantitative tightening of the neutrino mass limits, our analysis demonstrates that the hierarchy dependence of cosmological neutrino mass constraints remains remarkably stable when the small-scale CMB information is significantly improved by ACT DR6 and combined with high-precision BAO measurements from DESI DR2. This highlights the robustness of our conclusions against changes in both observational datasets and DE modeling.

In addition to the constraints derived from the baseline observational datasets discussed above, several recent studies have explored the total neutrino mass using a variety of other cosmological probes. 
For example, early massive galaxy candidates observed by the James Webb Space Telescope (JWST) have been shown to provide additional constraint power on the neutrino mass $\sum m_\nu$ in the $\Lambda$CDM, $w$CDM, and $w_0w_a$CDM models~\cite{Zhou:2025nkb}.
In Ref.~\cite{Feng:2024lzh}, further forecast shows that future gravitational-wave (GW) standard siren observations will improve the constraints on the total neutrino mass in interacting DE models. 
Furthermore, it is worth noting that Ref.~\cite{Du:2025xes} additionally incorporated weak lensing data and simultaneously considered $\sum m_\nu$ and the effective number of neutrino species $N_{\rm eff}$, thereby obtaining a non-zero neutrino mass at a $2.7\sigma$ level. In contrast, since this work does not consider additional lensing data and the $N_{\rm eff}$ parameter, we only obtained an upper limit on the total neutrino mass. Taken together, these studies indicate that the inclusion of JWST, GW, or lensing data can effectively enhance the cosmological constraints on $\sum m_\nu$. Moving forward, we will conduct in-depth investigations combining more comprehensive cosmological datasets to achieve more robust constraints on the neutrino mass.

In summary, using the latest ACT DR6 CMB and DESI DR2 BAO data, we find that the cosmological constraints on the total neutrino mass and its hierarchy remain robust across the dynamical DE models considered in this work. The improved small-scale CMB information provided by ACT DR6 not only leads to systematically tighter bounds on $\sum m_\nu$, but also confirms the stability of the hierarchy-dependent features of the neutrino mass constraints. These results highlight the critical role of high-resolution CMB observations, in combination with precision BAO measurements, in probing the absolute neutrino mass scale.

\section{Conclusion}\label{sec5}
In this paper, we employ the latest cosmological datasets, including ACT DR6 CMB, DESI DR2 BAO, and DESY5 SN measurements, to constrain the total neutrino mass and its hierarchy within the $\Lambda$CDM, $w$CDM, HDE, and $w_0w_a$CDM cosmological models. We aim to assess how the properties of DE and the improved small-scale CMB information provided by ACT DR6 jointly affect cosmological neutrino mass constraints in dynamical DE scenarios.

We find that the properties of DE can significantly influence the upper limits on the total neutrino mass $\sum m_\nu$. Compared with the $\Lambda$CDM model, the $w$CDM and HDE models yield more stringent constraints on $\sum m_\nu$, whereas the $w_0w_a$CDM model leads to substantially weaker bounds. Despite these model-dependent variations in the absolute limits, the hierarchy dependence of the neutrino mass constraints remains robust across all DE models, with the IH consistently yielding the loosest constraints and the DH providing the tightest ones.

By replacing the Planck CMB and DESI DR1 BAO data with the latest ACT DR6 and DESI DR2 measurements, we find that the improved small-scale CMB information from ACT DR6, combined with high-precision BAO observations from DESI DR2, systematically tightens the constraints on $\sum m_\nu$ across all neutrino mass hierarchies and DE models. More importantly, our results indicate that the hierarchy dependence of cosmological neutrino mass constraints remains remarkably robust, even when the quality of the CMB data is substantially improved beyond that of Planck. This highlights the crucial role of precise small-scale CMB measurements in breaking degeneracies between neutrino and DE parameters and underscores the robustness of cosmological neutrino mass constraints against changes in both observational datasets and DE models.

\begin{acknowledgments}
We thank Sheng-Han Zhou for helpful discussions.
This work was supported by the National Natural Science Foundation of China (Grant Nos. 12305069, 12533001, 12575049, and 12473001),
the National SKA Program of China (Grants Nos. 2022SKA0110200 and 2022SKA0110203),
the China Manned Space Program (Grant No. CMS-CSST-2025-A02),
and the National 111 Project (Grant No.B16009).

\end{acknowledgments}

\bibliography{DEneutrino}

\end{document}